\documentstyle[preprint,aps]{revtex}

\tighten

\begin{document}

\draft

\title{Stripe Correlations of Spins and Holes and Phonon Heat Transport
       in Doped $\rm {\bf La_2CuO_4}$}

\author{O.~Baberski, A.~Lang, O.~Maldonado, M. H\"ucker,
        B.~B\"uchner}

\address{II. Physikalisches Institut, Universit\"at zu K\"oln,
         50937 K\"oln, Germany}

\author{A.~Freimuth}

\address{Physikalisches Institut, Universit\"at Karlsruhe,
         76128 Karlsruhe, Germany}

\date{\today}
\maketitle

\begin{abstract}
We present experimental evidence for a dramatic suppression of
the phononic thermal conductivity of rare earth and
Sr doped $\rm La_2CuO_4$. Remarkably, this suppression correlates with
the occurrence of superconductivity. Conventional models for the phonon heat
transport fail to explain these results. In contrast,
a straightforward explanation is possible in terms of static and
dynamic stripe correlations of holes and spins.
\end{abstract}

\vskip1cm


The structural phase transition from the orthorhombic
(LTO) to the low temperature tetragonal (LTT) phase observed in rare earth (RE) doped
$\rm La_{2-x}Sr_xCuO_4$ (LSCO) has a pronounced influence on the electronic
properties of these materials. In particular, in a certain composition range
the LTT phase is not superconducting, but antiferromagnetic order
occurs at low temperature and at finite Sr, i.e. charge
carrier concentration~\cite{Buchner1}-\cite{Tranquada}.
Various mechanisms have been suggested to explain the strong
sensitivity of the electronic structure to the subtle structural
changes associated with the phase transition as e.g. band structure
effects~\cite{Pickett}, hole concentration dependent commensurability
effects and charge density wave--like instabilities~\cite{Barisic}, as
well as a novel coupling of the charge carrier motion to the tilt
displacements of the $\rm CuO_6$ octahedra via
spin--orbit scattering~\cite{Bonesteel}.
More recently Tranquada {\it et al.}~\cite{Tranquada} have presented evidence
for an important role of stripe correlations of spins and holes,
i.e. an ordering in the $\rm CuO_2$ planes into antiphase antiferromagnetic
stripe domains separated by domain walls to which the holes
segregate.

We show in this letter that in addition to its well established
influence on the electronic properties the structural transition has
also dramatic consequences for the lattice dynamics. Our main result
is that the phononic contribution $\kappa_{ph}$ to the thermal conductivity
increases strongly below the structural transition, {\em but only if the
LTT phase is not superconducting}~\cite{supercon}.
Remarkably, the behavior of the
thermal conductivity of Sr doped $\rm La_2CuO_4$ without additional
RE doping suggests the even more general conclusion
that $\kappa_{ph}$ is strongly suppressed for {\em all superconducting}
compositions of RE  and Sr doped $\rm La_2CuO_4$.
Our results have strong implications for the interpretation
of heat transport in doped $\rm La_2CuO_4$. In particular,
the conventional models for phononic thermal conductivity based on
enhanced phonon--defect scattering on alloying or conventional
phonon--electron scattering fail to account for these observations.
In contrast, a straightforward
interpretation is possible in terms of static and dynamic stripe
correlations of spins and holes. If this interpretation is correct
our data in turn imply that stripe correlations are dynamic for
superconducting compositions, whereas they are either
static or absent for non--superconducting compositions~\cite{supercon}.

The samples used in this study have been prepared by the usual solid state
reaction~\cite{Breuer2}. They are well characterized with various
physical properties as described in
ref.~\cite{Buchner2,Buchner3,Buchner4,Breuer2}. The thermal
conductivity $\kappa$ was measured by a standard method
for compounds with a wide range of Sr  und RE concentrations, $x$ and
$y$, respectively. We note that the absolute values of the
thermal conductivity in sintered materials differ from
those in single crystals due to scattering from grain boundaries.
In contrast, the temperature and doping dependencies of $\kappa$ found
in single crystals and polycrystals of LSCO agree well with each
other~\cite{Nakamura2,Sera}. Since we shall analyse only the temperature and doping
dependence of $\kappa$ we show  normalized data throughout this paper.
We mention that the absolute
values of $\kappa$ at high temperatures do not change significantly
as a function of RE doping within experimental errors.

As an example for the typical behavior of the thermal conductivity of
RE doped LSCO we show in fig.~1 our results for a
Pr and a Nd doped sample both with a Sr content of
$x = 0.12$. The Pr doped sample does not show a
low temperature structural transition~\cite{Schafer}. Its thermal conductivity
decreases monotonously with
decreasing temperature, similar to what is found in
LSCO at a comparable Sr content but without additional
RE doping~\cite{Nakamura2,Sera} (inset fig.~1).
In the Nd doped sample the  structural transition to the LTT phase
at $T_{LT} \approx 80$K has a dramatic influence on the thermal
conductivity: $\kappa$ increases strongly below $T_{LT}$ reaching a
maximum around 25K. We note that this temperature
dependence of $\kappa$ in the LTT phase
is similar to that found for the purely phononic thermal conductivity of
undoped insulating $\rm La_2CuO_4$ (inset fig.~1).

In the high $T_c$ superconductors the thermal conductivity has
an electronic and a phononic contribution, $\kappa_{el}$ and $\kappa_{ph}$,
respectively~\cite{Uher}. Nevertheless, in the present case the
increase of $\kappa$ below $T_{LT}$ is clearly due to an increase
of $\kappa_{ph}$: Firstly, the electrical conductivity decreases below
$T_{LT}$~\cite{Buchner2,Buchner4}. According to the
Wiedemann--Franz law $\kappa_{el}$ should then
decrease also and the increase of $\kappa = \kappa_{ph} + \kappa_{el}$ must be due to an increase
of $\kappa_{ph}$. Secondly, the pronounced increase of $\kappa$
below $T_{LT}$ occurs also for strongly underdoped samples (see
fig.~2), where the charge carrier concentration is so low that
$\kappa_{el} \ll  \kappa_{ph}$ and any increase of $\kappa$ must be due to an increase of
$\kappa_{ph}$.

From the data presented so far one may conclude that it is the structural phase
transition which causes the strong increase of $\kappa_{ph}$ below
$T_{LT}$. However, this is not true in general:
We show in fig.~2 the thermal conductivity for Eu doped samples
($y=0.15$) with Sr contents between $0.05$ and $0.22$. X--ray
diffraction experiments reveal the occurrence of the low temperature
structural transition to the LTT phase at $T_{LT} \approx 120$K
in {\em all} these
samples~\cite{Buchner4}. However, the low temperature increase of $\kappa$
occurs only in the limited
Sr concentration range of $x\le 0.17$. Remarkably,
Eu doped LSCO with $y =0.15$ is not superconducting for
$x \le 0.17$, whereas for $x>0.17$ superconductivity
occurs~\cite{Buchner4}.

We find similar results in the Nd doped samples. As a measure of
the increase of $\kappa$ at low temperatures one may use the
magnitude $\Delta \kappa$ of the jump of $\kappa$ at $T_{LT}$, since it is
apparent from the data of fig.~2 that $\Delta \kappa $ correlates with the size
of the low temperature maximum. In fig.~3 we show $\Delta \kappa $
as a function of the Sr concentration for
samples with Nd contents of $y=0.3$ and
$y=0.6$. The low temperature increase of $\kappa$,
i.e. $\Delta \kappa > 0$, occur for
$x \le 0.17$ for Nd doping of $y=0.3$ and for $x \le 0.23$ for Nd doping
of $y=0.6$. A comparison of these concentration pairs to the low
temperature Nd/Sr phase diagram of ref.~\cite{Buchner2} shows
that they lie on the boundary separating the superconducting
and the non--superconducting region of the
LTT phase. Thus, the low temperature maximum of
$\kappa_{ph}$ occurs in the LTT phase only for
compositions which are not superconducting.

Taking together these results we arrive at a remarkable correlation of the
lattice dynamics with the electronic properties of the LTT-phase:
The phononic thermal conductivity of RE doped
$\rm La_{2-x}Sr_xCuO_4$ increases strongly below the low temperature
structural transition, but only if the LTT phase is not superconducting.

At this point it is instructive to reexamine the thermal conductivity of LSCO
without additional RE doping (inset fig.~1)~\cite{Nakamura2}.
Insulating $\rm La_2CuO_4$ has a
purely phononic thermal conductivity with a pronounced maximum
occurring around 25K. Notably, in strongly overdoped,
non--superconducting samples the maximum of $\kappa$ at low temperatures
occurs also. We shall come back to this below.
For intermediate Sr doping of $ x \le 0.25$, when
samples are superconducting, $\kappa$ is strongly suppressed and
the low temperature maximum is absent.
Obviously this Sr concentration dependence
of $\kappa$ in LSCO fits very well into the correlation between the magnitude
of $\kappa_{ph}$ and the occurrence of superconductivity suggested by our
results on the RE doped samples. We are thus lead
to the more general conclusion: Whenever Sr and
RE doped $\rm La_2CuO_4$ is superconducting,
the phononic thermal conductivity is strongly suppressed; in
particular, the low temperature maximum of $\kappa_{ph}$ is
absent then.

We shall now discuss this result within the conventional models
for phonon heat transport in high $T_c$ materials. Since $\kappa_{ph}$
depends strongly on the RE  and the Sr concentration the
scattering of the phonons should also depend on the doping.
Two possible mechanisms are then apparent: 1. Doping induced
phonon--impurity scattering and 2. phonon--electron scattering.
We note first that the suppression of $\kappa_{ph}$ with increasing
Sr doping in LSCO without additional RE doping (inset fig.1) is usually
attributed to phonon--defect/disorder scattering which increases upon
alloying~\cite{Nakamura2,Sera,Uher}. However, our data immediately rule out such an explanation
due to the reappearance of the maximum of $\kappa_{ph}$ at {\em large
Sr  and RE doping} in the non--superconducting LTT phase (fig.1).
Regarding conventional phonon--electron scattering we note that,
firstly, the suppression of $\kappa_{ph}$ does occur for
Sr concentrations of $x \simeq 0.05$. For such low charge carrier
concentrations phonon--electron scattering is unimportant.
Secondly, if phonon--electron scattering suppresses $\kappa_{ph}$
in the superconducting samples one expects that at least part of the
phononic thermal conductivity reappears below the
superconducting transition (which is in the same temperature
range as $T_{LT}$), since
the opening of the energy gap would strongly suppress phonon--electron
scattering. However, no such enhancement of $\kappa_{ph}$ is
found in LSCO or RE doped LSCO below $T_c$.

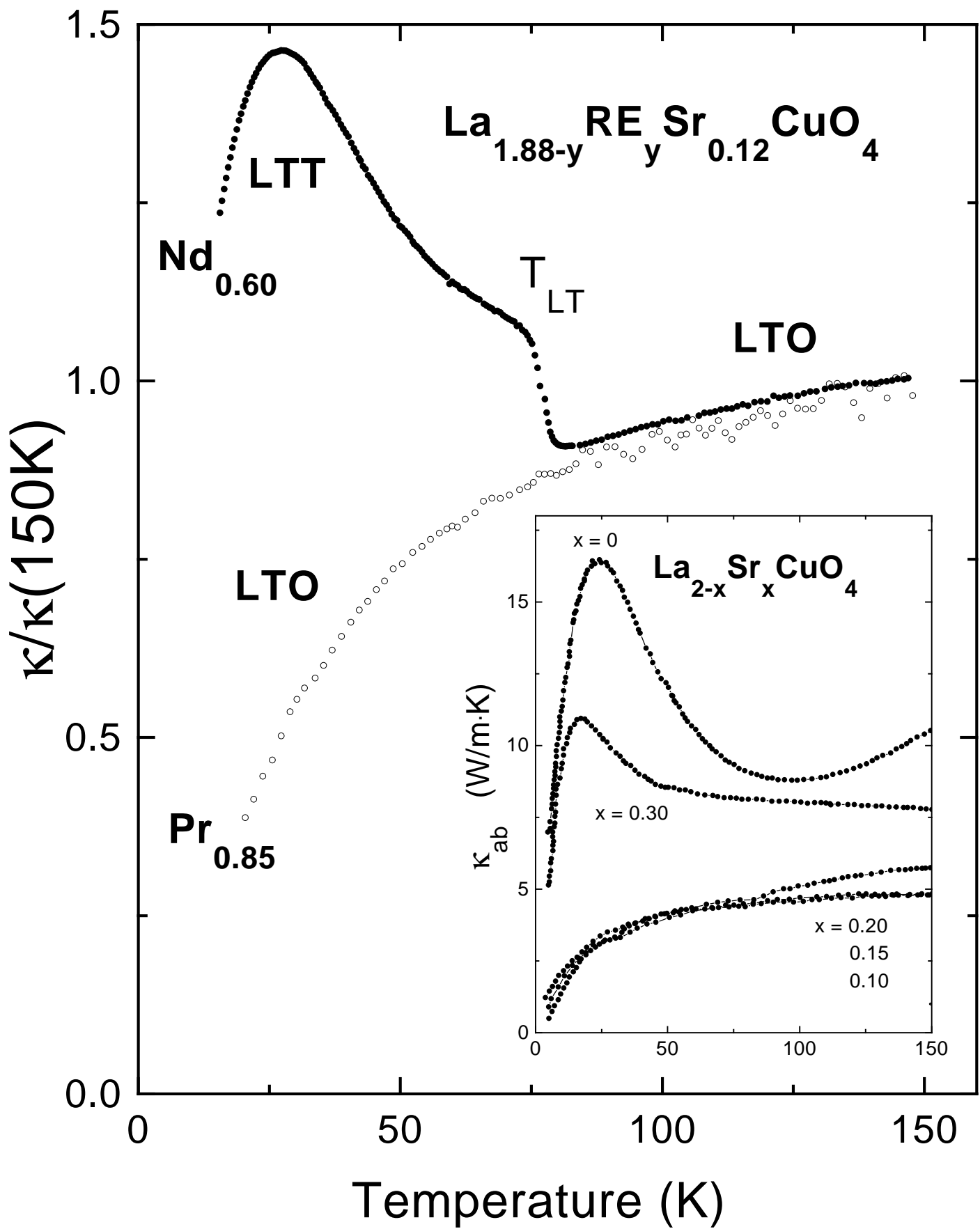
\begin{figure}
\caption[]{Thermal conductivity of Pr ($y=0.85$) and Nd ($y=0.6$) doped $\rm
La_{1.88-y}RE_ySr_{0.12}CuO_4$ normalized to $\kappa$(150K) as a function
of temperature (see text). Inset: In--plane thermal conductivity
$\kappa_{ab}$ of $\rm La_{2-x}Sr_xCuO_4$ single crystals as
a function of temperature taken from ref.~\cite{Nakamura2}.
Samples with $x=0.10, 0.15, 0.20$ are superconducting, those with
$x=0, 0.30$ are not.}
\label{fig1}
\end{figure}

We should mention here as a further mechanism a possible suppression of
$\kappa_{ph}$ due to scattering of phonons on anharmonic
soft phonon vibrations
associated with the structural transitions in doped $\rm La_2CuO_4$.
It is well known that the LTO phase is characterized by substantial
anharmonicity of the tilting vibrations of the $\rm CuO_6$-octahedra,
which seems to increase with Sr-doping~\cite{Braden}.
A possible reason for this anharmonicity is the
instability of the LTO-phase towards the
LTO $\rightarrow $ LTT phase transition.
However, it is well known that this structural instability
is mainly determined by sterical lattice properties which
depend on the RE content. In particular,
the low temperature transition does occur at comparable $T_{LT}$ for both
Sr doped compounds and insulators with
$x=0$~\cite{Buchner1,Buchner4,Keimer}.
It is apparent that the associated weak
Sr concentration dependence of the structural
instability does not correlate
with the pronounced change of $\kappa$ as a function of $x$ in
LSCO (inset of fig.~\ref{fig1}).
Moreover, a scenario which
explains the pronounced damping
of $\kappa_{ph}$ in LSCO due to the
structural instability becomes very unlikely
when taking into account the findings for insulators with $x$ = 0.
We find essentially the same $\kappa_{ph}$ in the entire
temperature range for $\rm La_2CuO_4$
and for RE doped compounds regardless whether the structure is
LTO or LTT. This means that neither the low temperature structure,
i.e. the presence of the tilting
instability at low temperatures in $\rm La_2CuO_4$ and
its absence in the LTT phase of the RE doped compounds,
nor the very close proximity to the structural
transition, which is apparentely
present close to $T_{LT}$ in the RE doped compounds,
causes significant differences
in the phonon heat transport of these
antiferromagnetic insulators.
Thus we conclude that anharmonicity due to the
structural instability can not explain
the damping of $\kappa_{ph}$ in LSCO ($x \neq 0$) and its
reappearance in the non--superconducting LTT phase. 
Note that this conclusion is
also supported by our findings for compounds with high Sr contents,
since the strong low temperature increase of $\kappa $ is
not linked to the phase transition alone
but to the electronic properties of the
LTT-phase (see fig.~2,3).
In particular, in the superconducting composition range of the
LTT-phase no low temperature enhancement of $\kappa_{ph}$ is observed,
although anharmonicity due to the structural instability
is absent.

We are thus led to the conclusion that conventional models for the
phonon heat transport fail to explain the experimental results for
RE and Sr doped $\rm La_2CuO_4$. On the other hand, the correlation
of the behavior of  $\kappa_{ph}$ at low temperatures with the occurrence of
superconductivity suggests an electronic scattering channel for the
phonons. If conventional phonon--electron scattering is inadequate, as
shown above, one might imagine scattering of phonons on some
`collective electronic excitation'. In the following we suggest
a possible such mechanism based on the formation of
so called stripe correlations of spins and holes and their dynamics.

The neutron scattering experiments of Tranquada et
al.~\cite{Tranquada} give evidence for {\em static} stripe
correlations, i.e. a static spatial modulation of the charge and spin
density in the $\rm CuO_2$ planes, in the {\em non--superconducting}
low temperature tetragonal phase of Nd doped LSCO ($y=0.4, x=0.12$).
Remarkably, the corresponding magnetic superstructure reflections occur exactly at the
wave vector of the well known incommensurate peaks in {\em inelastic} neutron
scattering on {\em superconducting} LSCO~\cite{Mason} (without additional
RE doping) leading to the suggestion that {\em dynamic}
stripe correlations are present in superconducting
LSCO~\cite{Tranquada}. 
Within this scenario a rather natural explanation of our results
is possible. We first note that via the
well known relation between bondlengths and charge density a spatially
inhomogeneous charge distribution implies corresponding variations of
the lattice constants. Note that just these variations are observed in
neutron scattering as superstructure reflections in the case of pinned
hole stripes in the non--superconducting LTT phase. If the
stripe correlations are dynamic one expects that dynamic lattice
modulations are induced provided that the time scale of the dynamic
stripe correlations is comparable to that of the lattice vibrations.
We note that, being a collective excitation linked to the magnetic
correlations, the time scale of the stripe correlations
is not given by the Fermi energy but should be much smaller.
In fact, the inelastic neutron results on superconducting Sr doped $\rm
La_2CuO_4$~\cite{Mason} suggest an energy scale for the dynamic stripe correlations
around 10meV, comparable with typical phonon energies.  Then the dynamics
of the stripe correlations will couple strongly to the lattice causing
a pronounced damping of the phonons~\cite{damping}.
Accordingly, one expects a strong  reduction of the lattice heat
conductivity. In contrast,
static stripe correlations will at most lead to some minor modifications
of the phonon dispersion compared to pure $\rm La_2CuO_4$, but they
will not suppress the phonon heat conductivity significantly.

Note that, if we assume that this scenario is the correct explanation
for the behavior of $\kappa_{ph}$, we may in turn
conclude from the correlation between the suppression of
$\kappa_{ph}$ and the occurrence of superconductivity that
the dynamics of the stripe correlations is important for
superconductivity.

Up to now we have discussed the behavior of $\kappa_{ph}$ at the structural
transition and have attributed the increase of $\kappa_{ph}$ to the {\em
pinning} of stripe correlations. However, we emphasize that the
reappearance of the maximum of $\kappa$ found in
strongly overdoped, non--superconducting LSCO without additional
RE doping (inset fig.~1) finds a straightforward
explanation within this picture also. At very high doping the
magnetic correlations in the $\rm CuO_2$ planes are weak or absent.
Therefore the tendency to form stripe correlations vanishes or is at
least strongly reduced.
Accordingly, the strong suppression of phonon heat conductivity discussed
above is absent and the phonon maximum should
reappear. We should note here that the maximum of $\kappa $ at high
Sr-doping is usually attributed to the electronic contribution to $\kappa $.
However, at such high Sr-doping disorder scattering should suppress any low
temperature maximum of the electronic contribution to the thermal conductivity,
as is well known for conventional metals. Therefore, be believe that
an electronic origin of the low temperature peak at high doping is
possible in principle, but rather unlikely.

We mention at this point that as a further check for our interpretation we have
measured the thermal conductivity of $\rm La_{2-x}Sr_xNiO_4$ with
$x=1/3$. This material is an insulator, i.e. $\kappa = \kappa_{ph}$.
Moreover, no structural transition occurs below 300K.
On the other hand, the presence
of charge and spin ordering below about 240K is well established in
this compound from neutron and electron diffraction
studies~\cite{nickel}. Remarkably, the behavior of $\kappa $
compares well with our findings for doped
$\rm La_2CuO_4$ in the non-superconducting
LTT-phase~\cite{kappanickel}.
Due to charge ordering $\kappa$ increases,
the slope $\partial \kappa/\partial T$ changes
sign from negative to positive, and at
low temperatures a pronounced maximum
of $\kappa$ occurs. These findings clearly confirm
the interpretation for the cuprates presented here.

We finally mention that when assuming a relation between the dynamics
of stripe correlations and superconductivity the influence of the
buckling distortion on the superconducting properties of the
LTT phase is qualitatively expected. We recall that it has been shown
in ref.~\cite{Buchner2} that the LTT phase is not superconducting, if
the buckling of the $CuO_2$ planes, i.e. the tilting of the
$CuO_6$ octahedra, exceeds a critical value~\cite{supercon}.
According to the results
of Tranquada {\it et al.} the pinned stripes in a single $\rm CuO_2$ layer
are either parallel to the  [100]-- or to the [010]--direction (using the
notation of an undistorted lattice). From the structure of the
LTT phase it is apparent that there are differences between these
directions, which increase with increasing buckling distortion.
Therefore, assuming that the pinning of the stripe correlations
requires a finite `pinning potential' one expects that within the
LTT phase stripe correlations are pinned only beyond a certain value
of the buckling.

In conclusion, we have shown that in RE  and Sr doped $\rm La_2CuO_4$
the phononic thermal conductivity is strongly suppressed for {\em all}
superconducting compositions. The conventional models of phonon heat transport based on
phonon--defect scattering or conventional phonon--electron scattering
fail to explain these results. In contrast, the recently
suggested picture of dynamic and static stripe correlations of spins
and holes allows for a straightforward explanation. If this
explanation is correct the correlation between suppressed $\kappa_{ph}$ and the
occurrence of superconductivity tells that the dynamics of stripe
correlations is important for superconductivity in doped LSCO.

\begin{acknowledgements}
This work was supported by the Deutsche Forschungsgemeinschaft through SFB 341.
We thank W.~Brenig, C.~Bruder, P. Hirschfeld, and A.P.~Kampf for useful
discussions.
\end{acknowledgements}

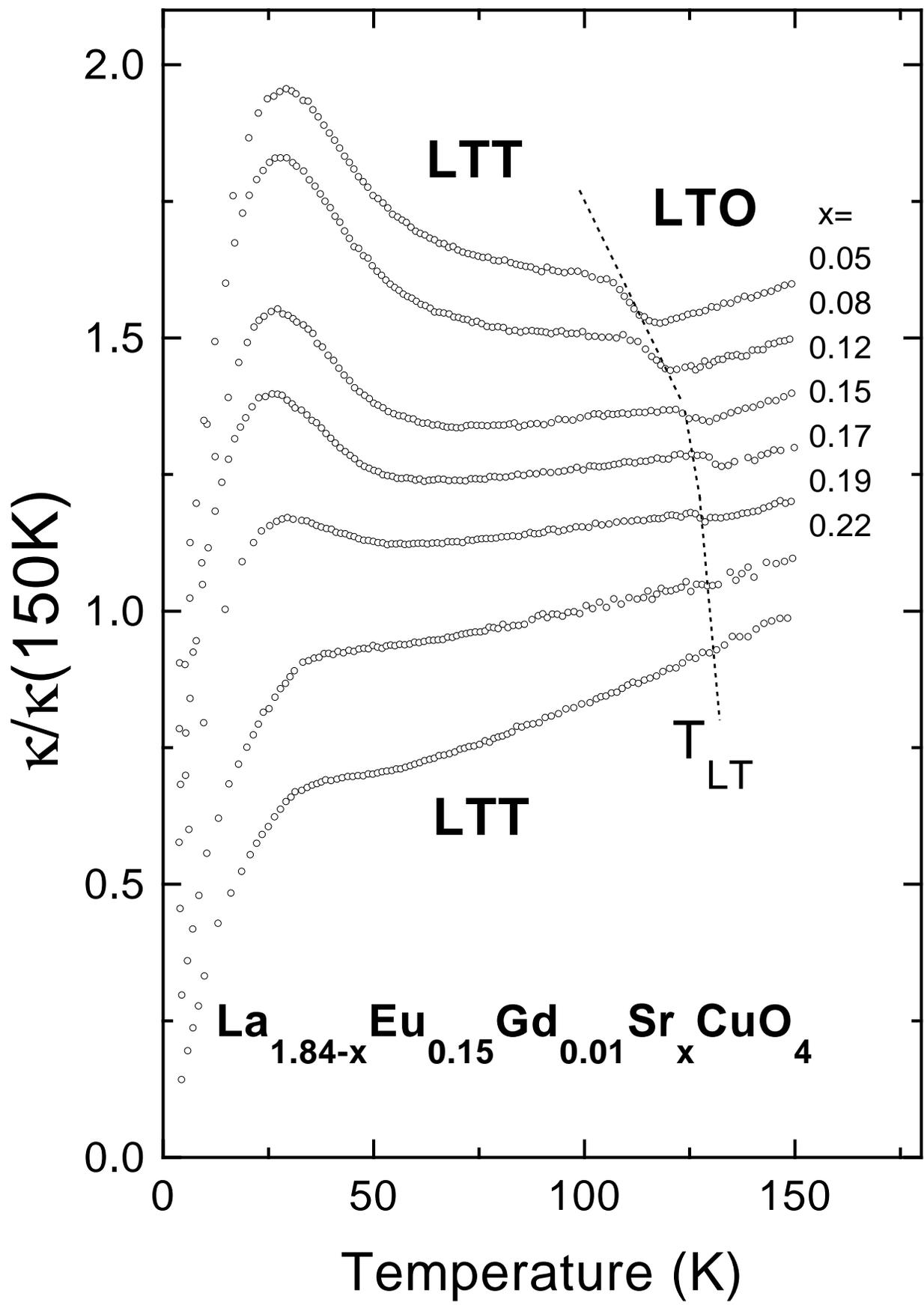
\begin{figure}
\caption{Thermal conductivity  of Eu doped
$\rm La_{1.84-x}Eu_{0.15}Gd_{0.01}Sr_xCuO_4$ normalized to $\kappa$(150K) as a
function of temperature for various Sr concentrations given in the figure.
Curves are shifted for clarity. Only samples with $x>0.17$ are
superconducting. $T_{LT} \approx 120$K is indicated by
the dashed line. The small amount of Gd in the samples was used as an
ESR-probe and is unimportant in the context of this paper.}
\label{fig2}
\end{figure}

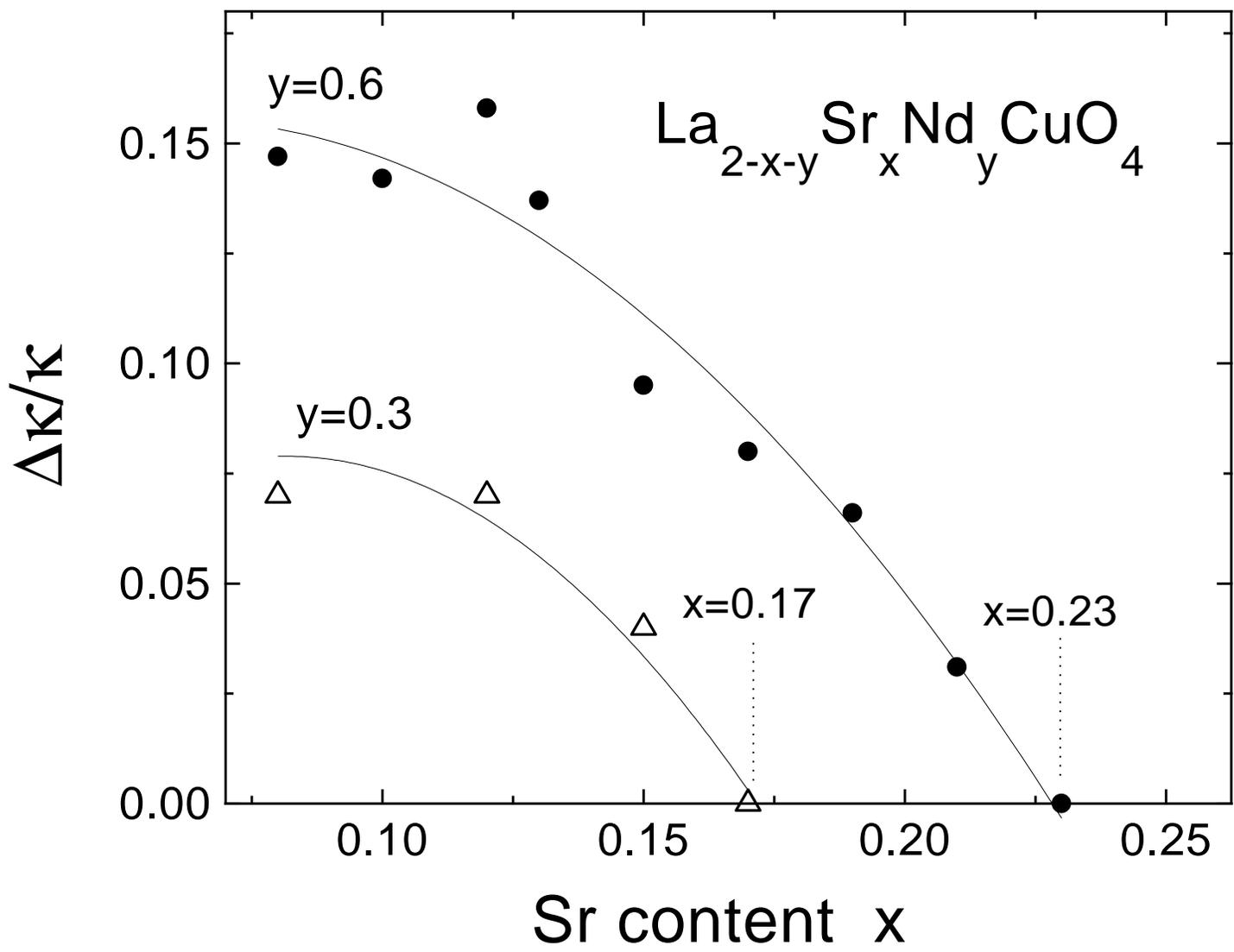
\begin{figure}
\caption{Anomaly $\Delta \kappa$ extracted from the jump--like anomalies of
$\kappa$ at $T_{LT}$ (see fig.1) normalized to $\kappa(T_{LT})$ as a function
of the Sr concentration for Nd doped $\rm La_{2-x}Sr_xCuO_4$.
$\bigtriangleup$ : Nd content $y=0.3$; $\bullet  : y =0.6$.}
\label{fig3}
\end{figure}

\end{document}